\documentclass[modern, onecolumn]{aastex631} 
\usepackage{multirow}
\usepackage{booktabs} 
\usepackage{commath}

\newcommand\kms{km s$^{-1}$}

\begin{document}

\title{A Galactic Supernova Remnant Candidate at \textit{l}=25${\fdg}$8, \textit{b}=+0${\fdg}$2 Revealed by Near-Infrared Imaging and Spectroscopy}

\author[0000-0002-5015-8223]{Yesol Kim}
\affiliation{Korea Astronomy and Space Science Institute, 776, Daedeok-daero, Yuseong-gu, Daejeon, 34055, Republic of Korea}
\affiliation{Department of Physics and Astronomy, Seoul National University, 1 Gwanak-ro, Gwanak-gu, Seoul 08826, Republic of Korea}
\author[0000-0002-2755-1879]{Bon-Chul Koo}
\affiliation{Department of Physics and Astronomy, Seoul National University, 1 Gwanak-ro, Gwanak-gu, Seoul 08826, Republic of Korea}
\author[0000-0003-0894-7824]{Jae-Joon Lee}
\affiliation{Korea Astronomy and Space Science Institute, 776, Daedeok-daero, Yuseong-gu, Daejeon, 34055, Republic of Korea}

\begin{abstract}
We present high-resolution near-infrared spectroscopic observations of the newly identified supernova remnant (SNR) candidate G25.8+0.2 obtained with the Immersion Grating Infrared Spectrograph (IGRINS) on the Harlan J. Smith Telescope. The source was originally discovered in the UKIRT Wide-field Infrared Survey for Fe$^{+}$ (UWIFE; Lee et al. 2014). Our spectra reveal multiple kinematic components in the [Fe II] 1.644 $\mu$m emission. The high-velocity components exhibit elevated [Fe II]/Br $\gamma$ ratios characteristic of shock excitation, while the low-velocity components are dominated by hydrogen and helium recombination lines and are consistent with photoionized gas, indicating an H II-region origin.
G25.8+0.2 lies within the G26 complex, a large ($\sim$15' × 30', corresponding to $\sim$28 pc × 57 pc at a distance of 6.5 kpc) star-forming region in the inner Galaxy. The shock-excited [Fe II] filaments closely trace the morphology of the bright radio shell that partially encloses centrally filled soft X-ray emission, strongly suggesting recent supernova activity in this region. We discuss the physical nature of G25.8+0.2 and its relationship to the surrounding G26 star-forming complex. In addition, we derive the extinction toward the source using Brackett-line ratios and constrain the spectral types of the dominant ionizing stars from the He I 2.058 $\mu$m/Br $\gamma$ line ratios. 
\end{abstract}


\section{Introduction} \label{sec:intro}

Supernova remnants (SNRs) are key dynamical drivers in the interstellar medium (ISM), injecting energy through shock waves that shape the surrounding interstellar and circumstellar media. At present, approximately 310 Galactic SNRs have been catalogued \citep{Green+25}. Most SNRs have been identified and characterized in the radio band, where non-thermal synchrotron emission contrasts with the thermal bremsstrahlung typically observed in H II regions \citep{Dubner+15}. However, recent advances in instrumentation and various search techniques, such as blind surveys, have led to an increase in the number of identified SNRs at other wavelengths (\citealt{Long+2017}, \citealt{Fesen+2024}).

With the recent advent of near-infrared (NIR) instruments, shock tracers in the NIR have been proposed as a new technique for identifying SNRs. In particular, NIR [Fe II] emission lines have been increasingly used to identify SNRs, especially in highly obscured regions where the majority of core-collapse supernovae (CCSNe) are born. Since an elevated [Fe II]/H ratio indicates a shock-excited origin of the [Fe II] emission --- analogous to the [S II]/H${\alpha}$ ratio in the optical --- the use of [Fe II] as a diagnostic tool for probing SNR populations has become more widespread. This approach has been applied to Galactic SNRs (e.g., G11.2\textminus0.3; \citealt{Koo+07}, 3C 396; \citealt{Lee+09}, IC 443; \citealt{Kokusho+13}, W49B; \citealt{Keohane+07}), as well as to several nearby galaxies, including M82 \citep{Greenhouse+91}, NGC 1569, NGC 3738, and NGC 5253 \citep{Labrie+06}, NGC 6946 (\citealt{Bruursema+14}; \citealt{Long+20}), M83 \citep{Blair+14}, and M33 \citep{Morel+2002}.  

The United Kingdom Infrared Telescope (UKIRT) Wide-field Infrared Survey for Fe$^{+}$ (UWIFE; \citealt{Lee+14}) is an unbiased Galactic survey of the [Fe II] a$^{4}$D$_{7/2}$ ${\rightarrow}$ a$^{4}$$F_{9/2}$ 1.644 ${\mu}$m emission line. This survey encompasses approximately 180 deg$^{2}$ of the first Galactic quadrant (7$^{\circ}$ ${\leq}$ $\mathit{l}$  ${\leq}$ 62$^{\circ}$, \abs{b} ${\sim}$ 1$\rlap.^{\circ}$5) using a narrow-band [Fe II] filter. The [Fe II] line preferentially traces dense shocked-gas, and the UWIFE data provide unprecedented depth and high spatial resolution (median seeing of 0.83", 5${\sigma}$ detection limit of 18.7 mag, and a surface brightness limit down to 8.1 ${\times}$ 10$^{-20}$ W m$^{-2}$ arcsec$^{-2}$). We designated extended [Fe II] 1.644 ${\mu}$m emission-line sources as Ionized Fe Objects (IFOs) and aimed to identify the entire population of extended IFOs within the UWIFE survey area \citep{Kim+24}. A subsequent compilation of known Galactic SNRs exhibiting [Fe II] emission in the UWIFE survey area was reported by \cite{Lee+19}. 

\begin{figure}
\begin{center}
\includegraphics[width=155mm]{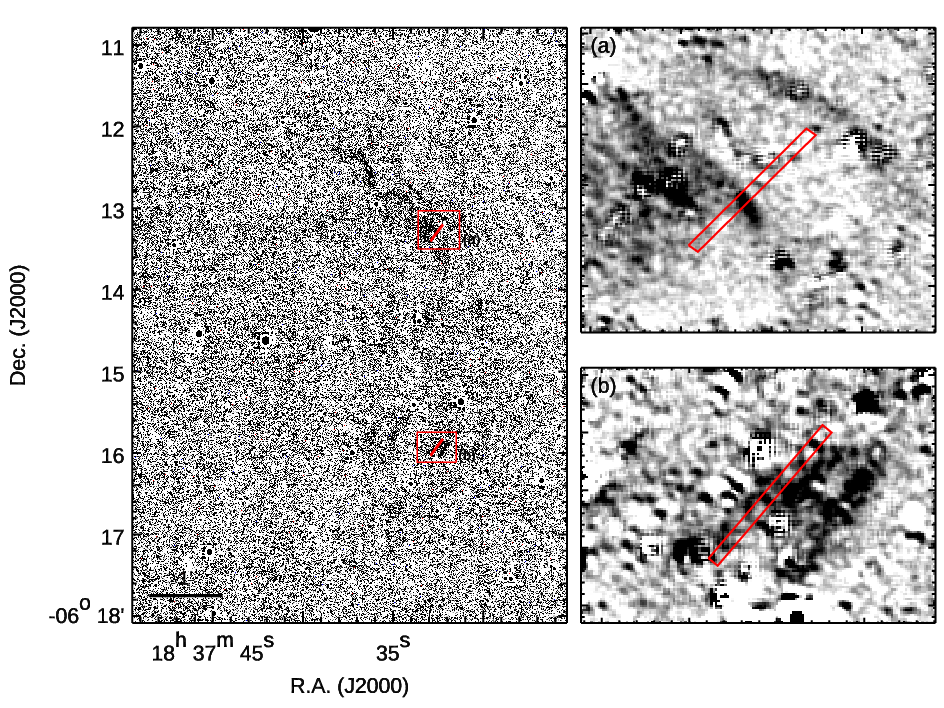} 
\end{center}
\caption{(\textit{left}) Continuum-subtracted [Fe II] 1.644 ${\mu}$m narrow-band image of G25.8+0.2 region. 
Red rectangles show slit positions.  (\textit{right}) Panel (a) and (b) are magnified image of the continuum-subtracted [Fe II] 1.644 ${\mu}$m image in left, showing the slit positions of our NIR spectroscopy.}
\label{fig:fe_2024}
\end{figure}

During the blind search for IFOs, we serendipitously discovered an partial shell-like IFO of unknown origin, designated IFO 96, located near \textit{l} ${\sim}$ 26${\arcdeg}$ at $\alpha = 18^{\mathrm{h}}37^{\mathrm{m}}40\fs83$, $\delta = -06^{\circ}14^{\prime}52^{\prime\prime}$ (J2000) (Figure \ref{fig:fe_2024}). Among the possible counterparts of IFOs - such as outflows of young stellar objects, H II regions, SNRs, planetary nebulae, and nebulae of luminous blue variables - the high flux (2.21 ${\times}$ 10$^{-15}$ W m$^{-2}$) and large angular extent ($\sim6^{\prime}$) of IFO 96 are comparable to those of H II regions or SNR-type IFOs. This discovery prompted a spectroscopic follow-up study to investigate the origin of IFO 96.

The region containing IFO~96 spatially coincides with a known radio continuum structure cataloged as G025.8+00.2 (hereafter, G25.8+0.2), which constitutes part of a larger U-shaped radio complex commonly referred to as the ``G26 complex." Hydrogen radio recombination lines have been detected toward this source \citep{Sewilo+04, Quireza+06_recom}, and the spectral index of the brightest portion of the radio emission was measured to be flat \citep{DAmico+01}, indicating that the radio emission is dominated by thermal free-free processes characteristic of an H II region. On the other hand, using fluxes compiled from published radio catalogs, \citet{Stein+21} reported that the system consists of two components with markedly different spectral indices, $\alpha = +1.07$ and $-0.56$, respectively. This result suggests that a non-thermal component may be present within the G26 system, although the compact source G25.8+0.2 itself is included in the master catalog of Galactic H II regions compiled by \citet{Paladini+03}.

In addition to the radio continuum emission, extended soft X-ray emission has been detected in the vicinity of G25.8+0.2 \citep{Sakamoto+01}. This X-ray source was serendipitously discovered during a search for the counterpart of an EGRET $\gamma$-ray source and exhibits a spectrum that can be fitted with an optically thin thermal plasma model, suggesting an origin related to a supernova remnant. The presence of nearby pulsars has also been noted \citep{DAmico+01, Lin+08}, although their physical association with the soft X-ray emission remains uncertain.

The most recent comprehensive investigation of the G26 complex was carried out by \citet{Cichowolski+18} (hereafter C18), who conducted a multiwavelength analysis of the region. They showed that G25.8+0.2 corresponds to the southwestern portion of a large U-shaped radio arc that is in direct contact with adjacent molecular clouds. Based on this morphology and the associated molecular gas distribution, C18 argued that the G26 complex represents an active star-forming region in the inner Galaxy at a distance of $\sim$6.5~kpc, in which the radio continuum emission arises primarily from photoionized gas produced by embedded massive stars. They identified several candidate O-type stars that could be responsible for the ionization. At the same time, C18 noted that one or more SN explosions could have contributed to the formation of the radio shell. The discovery of extended [Fe~II] emission associated with G25.8+0.2 adds a new dimension to this picture and the interpretation of the origin and evolutionary history of the G26 complex.

In this paper, we present a high-resolution NIR spectroscopic analysis of a large-scale, partial shell of [Fe II] emission newly identified in the UWIFE survey, based on \textit{H}- and \textit{K}-band observations. The contents of this paper are as follows. In Section 2, we describe the IGRINS spectroscopic observations and data reduction procedures. The results of the observations are presented in Section 3, and in Section 4, we discuss the physical origin of the [Fe II] emission in G25.8+0.2 and explore its relationship to the G26 complex using multiwavelength data.

\section{Observation and Data} \label{sec:data}
\subsection{NIR Spectroscopy and Data Reduction} 
We carried out NIR spectroscopic follow-up observations of IFO 96 using the Immersion Grating Infrared Spectrograph (IGRINS; \citealt{Yuk+10}) mounted on the 2.7 m Harlan J. Smith Telescope at McDonald Observatory, University of Texas at Austin. IGRINS is a cross-dispersed, high-resolution echelle spectrograph (${\lambda}$/${\Delta}$${\lambda}$ ${\equiv}$ R ${\sim}$45,000) that simultaneously covers the NIR \textit{H}- and \textit{K}-bands (1.49--1.81 and 1.93--2.46 ${\mu}$m). The corresponding velocity resolution is ${\Delta}$v = 7 km s$^{-1}$, with a ${\sim}$3.5 pixel sampling per resolution element \citep{Park+14}. The slit width and length at the telescope were 1${\arcsec}$ and 15${\arcsec}$, respectively. The observations were conducted on 2015 August 6 UT, during which we obtained spectra at two slit positions along the bright [Fe II] filaments (Figure~\ref{fig:fe_2024}). The orientation of slit 1 was set perpendicular to the northern filament, whereas that of slit 2 was almost parallel to the southern filament. We took two 340~s on-source and two 340~s off-source (sky) exposure, the latter located $15^{\prime}$ south of the target. For flux calibration, we also observed the nearby A0V star 65 Sgr (\textit{K} = 6.341 mag) with four exposures of 90 s each, immediately after the target, under similar weather and airmass conditions. A detailed log of the observations is presented in Table~\ref{table:log}. 

We utilised the Python-based data reduction pipeline IGRINS PLP (v2.2.0-alpha.1\footnote{https://github.com/igrins/plp};\citealt{Lee+17}) for the initial processing of the IGRINS data. The data were reduced using the \textit{extended on-off} mode, which is specifically designed for extended source reduction. The pipeline performs bad-pixel masking, sky subtraction, flat-fielding, and aperture extraction. Wavelength calibration was determined using the UrNe lamp frame obtained during the observations. For the two-dimensional spectra, each order was flattened, and the low-signal regions near the first and last rows of the echelle format were trimmed. Additional flux calibration was performed using a manually developed IDL routine designed for further processing of the two-dimensional spectra. This step included telluric correction and absolute flux calibration using a Kurucz model spectrum of an A0V star \citep{Rieke+08} with a resolving power of R=100,000\footnote{http://kurucz.harvard.edu/stars/vega/}, interpolated to match that of IGRINS. Because sky lines varied during the exposures, the subtraction of sky lines was not optimal, resulting in residuals (e.g., in the velocity range of ${\sim}$ 85--100 km s$^{-1}$ adjacent to the He I 2.058 ${\mu}$m line, see Figure ~\ref{fig:2d}). We masked these residual sky lines before further analysis. Since the observer's motion toward the target, computed using the IRAF task \texttt{rvcorrect}, did not exceed 1 km s$^{-1}$--much smaller than the instrumental velocity resolution of IGRINS--the velocity correction was not applied.

\subsection{Auxiliary Data}

We used radio continuum, X-ray, and $^{12}$CO ($J$=1--0) line survey data of the target region 
to investigate possible associations with other objects and their interactions.

\subsubsection{Radio}
We used the high-resolution 20 cm radio continuum data from the Multi-Array Galactic Plane Imaging Survey (MAGPIS; \citealt{Helfand+06}), which mapped part of the first Galactic quadrant (5$^{\circ}$ ${\textless}$ $\mathit{l}$ ${\textless}$ 48$^{\circ}$.5, \abs{b} ${\textless}$ 0.8$^{\circ}$). The data were obtained in the B, C, and D configurations of the Very Large Array (VLA) between March 2001 and March 2004. The angular resolution is ${\sim}$6${\arcsec}$ and the typical rms is 0.3 mJy. The missing flux from large-scale structures (${\gg}$ 1${\arcmin}$), to which the VLA is insensitive, was compensated by combining the data with the 1400 MHz survey conducted with the Effelsberg 100 m telescope \citep{Reich+90}, which has an angular resolution of ${\sim}$9.4${\arcmin}$. The reduced, full-resolution 20 cm images were downloaded as FITS files from the MAGPIS survey website\footnote{http://third.ucllnl.org/gps}.

\subsubsection{X-ray} 
We utilized X-ray observations of the region obtained with the Advanced Satellite for Cosmology and Astrophysics (\textit{ASCA}), which was equipped with the Gas Imaging Spectrometer (GIS) operating in both the soft (0.7 -- 2.0 keV) and the hard (4 -- 10 keV) bands. These observations were conducted as part of an X-ray follow-up of bright, unidentified Energetic Gamma Ray Experiment Telescope (EGRET) sources \citep{Roberts+01}. The observations were carried out on 1998 April 2 and October 18, covering the 95\% confidence contour of the ${\gamma}$-ray source 3EG J1837-0606. The processed, revision 2 public data were retrieved from the Data ARchives and Transmission System (DARTS)\footnote{https://data.darts.isas.jaxa.jp/pub/}, operated by the Center for Science-satellite Operation and Data Archive (C-SODA).

\subsubsection{$^{12}$CO J = 1--0}
We used the $^{12}$CO (\textit{J} = 1--0) data from the FOREST (FOur-beam REceiver System on the 45m Telescope) Unbiased Galactic Plane Imaging survey with the Nobeyama 45m telescope (FUGIN; \citealt{Umemoto+17}) to investigate the molecular gas distribution in the G26 region. The survey simultaneously observed the $^{12}$CO, $^{13}$CO, and C$^{18}$O J= 1--0 transitions. It covered portions of the first (10$^{\circ}$ ${\textless}$ $\mathit{l}$ ${\textless}$ 50$^{\circ}$, \abs{b} ${\textless}$ 1$^{\circ}$) and third (198$^{\circ}$ ${\textless}$ $\mathit{l}$ ${\textless}$ 236$^{\circ}$, \abs{b} ${\textless}$ 1$^{\circ}$) Galactic quadrants between 2014 April and 2017 March. For the $^{12}$CO (\textit{J} = 1--0) line, the angular resolution is 20${\arcsec}$ and the velocity resolution is 1.3 km s$^{-1}$. The expected rms (T$_{A}^{*}$) of the $^{12}$CO data is 0.24 K. The processed FITS cube data were retrieved from the Japanese Virtual Observatory (JVO) portal\footnote{http://jvo.nao.ac.jp/portal/nobeyama/fugin.do}.

\section{Results} \label{sec:resul}

\subsection{Spectra and line ratios} \label{sec:spec}
\begin{figure}[ht]
\begin{center} 
\includegraphics[width=45mm,angle=270]{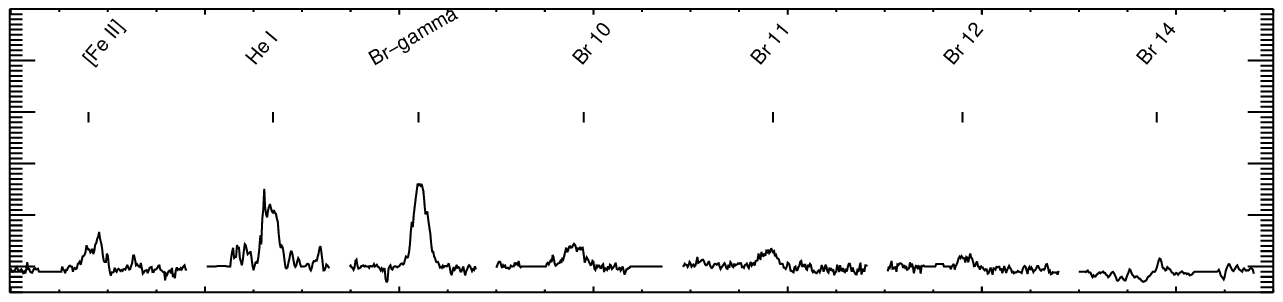} \\
\includegraphics[width=45mm,angle=270]{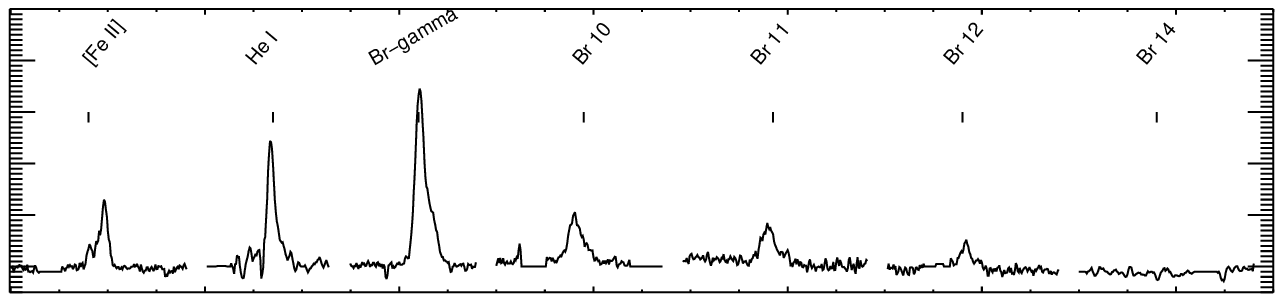}
\caption{1D spectra of identified emission lines. (Upper) slit 1; (lower) slit 2.}
\label{fig:1d}
\end{center}
\end{figure}

\begin{figure}
\center
\includegraphics[width=12cm]{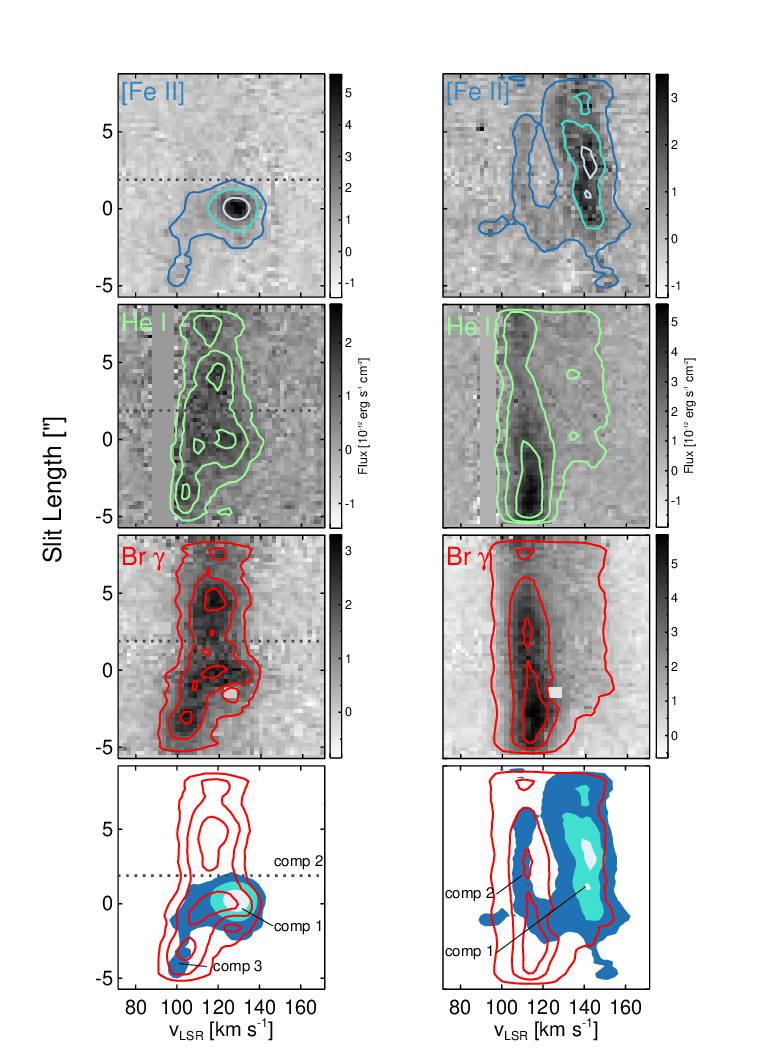}
\caption{Position-velocity (PV) diagrams of the detected emission lines in slit~1 (left) and slit~2 (right). The top three panels show the [Fe\,II] 1.644~$\mu$m, He\,I 2.058~$\mu$m, and Br $\gamma$ 2.166~$\mu$m emission lines. The flux scale of the color bar is in units of $10^{-19}$~W~m$^{-2}$. Velocity channels between 90 and 100~km~s$^{-1}$ in the He\,I panels, which are contaminated by OH residuals, are masked. The bottom panel presents contour maps of the [Fe\,II] (filled blue) and Br $\gamma$ (red) emission over the same position-velocity range.} 
\label{fig:2d}
\end{figure}

 Figure~\ref{fig:1d} shows one-dimensional (1D) spectra of the detected emission lines. In both slit positions, we identified [Fe II] 1.644 ${\mu}$m, He I 2.058 ${\mu}$m, Br ${\gamma}$ 2.166 ${\mu}$m, and Br 10\textminus14 lines. 
In the figure, one can see that the profiles of the [Fe II] line   
 differ from those of the H and He lines, suggesting distinct velocity structures. The peaks of the [Fe II] lines are at higher velocities than those of H and He lines. 
 This difference is more clearly illustrated in Figure~\ref{fig:2d}, which shows  
 position-velocity (PV) diagrams of the [Fe II], He I, and Br ${\gamma}$ lines. 
 The three lines occupy similar velocity ranges---v$_{LSR}$ ${\sim}$ 90\textminus140 and ${\sim}$ 100\textminus 160 km s$^{-1}$ in slits 1 and 2, respectively. 
 In slit 1, [Fe II] emission is almost absent in the upper part of slit, whereas 
 a compact and bright [Fe II] feature is present at ${\sim}$130 km s$^{-1}$ in the lower part of the slit.
 A faint [Fe II] emission is also seen at $\sim 100$~\kms, connected to the compact knot.
The position of this compact feature spatially coincides with the bright filament in the UWIFE [Fe II]-H image (Figure~\ref{fig:fe_2024}). 
The H and He I lines are present throughout the entire slit 1, and their velocity structures are similar to each other.
In the lower part of the slit, their velocity structures are also
similar to that of the [Fe II] emission.
At $y\simeq 0$ in Figure~\ref{fig:2d}, 
the line is broad with two velocity peaks;
the higher-velocity component is close to the compact [Fe II] emission.
We refer to the high-velocity component with strong [Fe II] emission
in the lower part of the slit as comp 1, 
the emission component bright in H and He lines in the upper slit area as comp 2, and the low-velocity component with faint [Fe II] emission 
in the lower part of the slit as comp 3.  

In slit 2, strong [Fe II] emission is detected at v$_{LSR}$ ${\sim}$+140~\kms\ along the slit, where the H and He emissions are very faint. 
Note that the orientation of slit 2 was intended to cover the [Fe II] structure in a parallel direction, whereas slit 1 intersects the structure perpendicularly. 
Bright H and He I lines, which exhibit similar velocity structures, 
appear at significantly lower velocities, around +110~\kms.
A faint [Fe II] emission component is also present at these velocities, spatially coincident with the H and He I lines.
We refer to the high-velocity component with strong [Fe II] emission as comp 1, and the low velocity component with strong H and He emission as comp 2. 

\begin{figure}[h]
\begin{center} 
\includegraphics[width=140mm,angle=0]{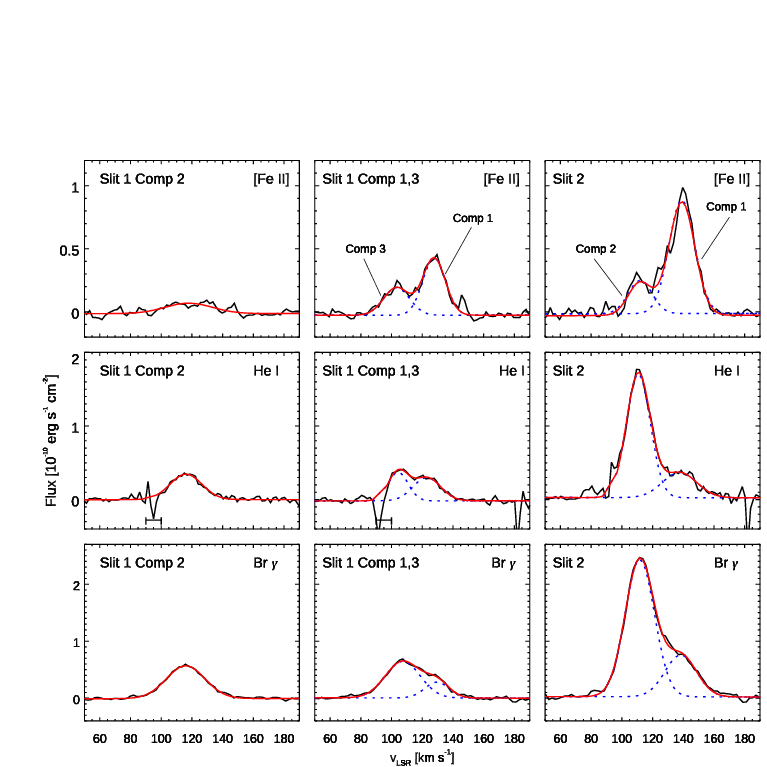} 
\caption{1D spectra of the [Fe II] 1.644 ${\mu}$m, He I 2.058 ${\mu}$m, and Br ${\gamma}$ 2.166 ${\mu}$m lines (from top to bottom) for the two slit positions. 
For slit 1, two spectra were extracted from the upper and lower sections of the slit, whereas for slit 2, a single spectrum was extracted from the entire slit (see Figure~\ref{fig:2d}). The black solid lines represent the observed spectra. The blue dotted lines show the individual Gaussian components, and the red solid lines indicate the combined model profiles obtained from the Gaussian fitting. When the spectra are fitted by two Gaussian components (Section 3.1), each component is indicated by a blue dotted line. The artifact at V$_{LSR}$ ${\sim}$ 90--100 km s$^{-1}$ in the He\,I spectrum of slit~1, marked with a horizontal bar with vertical end caps, was masked during the Gaussian fitting.} 
\label{fig:1d_fit}
\end{center}
\end{figure}

The 2D spectrum was collapsed into 1D spectra, and multi-component Gaussian fitting was performed to determine the line parameters of individual velocity components (Figure~\ref{fig:1d_fit}). 
For slit 1, two 1D spectra were extracted from the upper and lower sections of the slit, as indicated by the dotted lines in Figure~\ref{fig:2d}. 
In the lower section, where comp
1 and comp 3 are present, we applied a two-component Gaussian fit, whereas in the upper section (comp 2), a single component fit was used.
For slit 2, a single 1D spectrum was extracted from the entire slit and modeled with a two-component Gaussian fit.
The fitting was performed using the IDL routine \texttt{XGAUSSFIT}. 
For the [Fe II] and He I lines, the central velocity, peak intensity, and line width were treated as free parameters during the fitting process.
For the Br ${\gamma}$ lines, the line widths were fixed to those of the fitted [Fe II] lines, while the remaining parameters were allowed to vary. 
The results of the fitting are summarized in Table~\ref{table:mecha}. 

Table~\ref{table:mecha} shows that the line ratio [Fe II] 1.644 ${\mu}$m/Br ${\gamma}$ (hereafter R([Fe II]/Br ${\gamma}$) of comp 1 in slit 1 is greater than unity, whereas those of the other components in the same slit are $<0.2$. A similar trend is observed in slit 2, where R([Fe II]/Br ${\gamma}$) for comp 1 is 0.76 --- over an order of magnitude higher than that of comp 2 (0.07). The extinction-corrected ratios show an even greater contrast (see Section \ref{sec:ext}). In addition, these components also have velocities distinct from the others: comp 1 in slit 1 is redshifted by several tens of km s$^{-1}$ relative to the other components in the same slit, as is comp 1 in slit 2. The origin of this emission will be discussed in more detail in Section \ref{sec:dis}.

\subsection{Extinction} \label{sec:ext}
\begin{figure}
\center
\includegraphics[width=13cm]{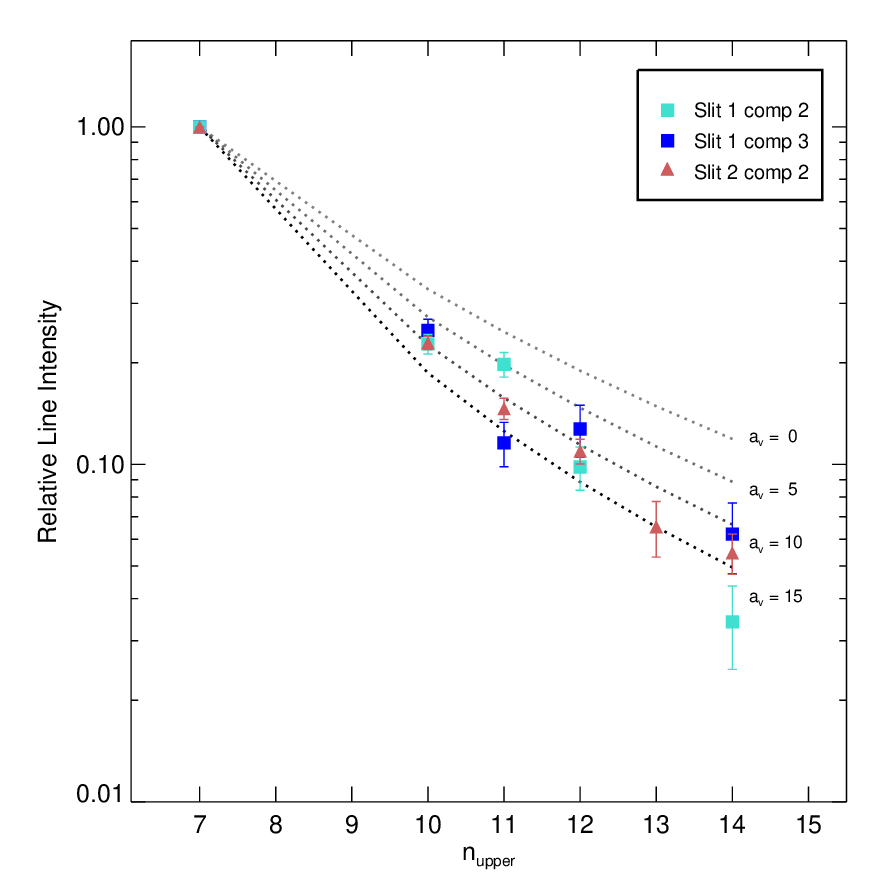}
\caption{Observed Brackett-line intensities relative to Br$\gamma$ in the IGRINS spectra of the two slit positions. The dotted lines show the Case~B theoretical predictions for $T_{e}=10^{4}$~K and $n_{e}=10^{4}$~cm$^{-3}$ \citep{Hummer+87}, shown for a range of extinctions.} 
\label{fig:extin}
\end{figure}

Extinction toward the two slit positions of G25.8+0.2 was measured using the flux ratios among the Brackett series lines. 
Specifically, the flux of each Brackett line was divided by that of Br ${\gamma}$, and the resulting line ratios are listed in Table~\ref{table:mecha}. 
The uncertainties in these ratios were derived using standard error propagation, accounting for the individual flux uncertainties of each line.  
To estimate the extinction, the observed Brackett line ratios were compared with their theoretical values predicted by recombination theory under optically thin conditions. 
According to Case B recombination, the strengths of the higher-order Brackett lines follow a predictable sequence of decreasing intensity. 
For this comparison, we focused on comp 2 and 3 in slit 1 and comp 2 of slit 2, which are likely associated with photoionized gas. 
The theoretical Brackett line ratios were adopted from \citet{Hummer+87}, assuming an electron temperature of T$_{e}$ = 10$^{4}$ K and an electron density of n$_{e}$ = 10$^{4}$ cm$^{-3}$. 
We note that lowering the density to n$_{e}$ = 10$^{2}$ cm$^{-3}$ yields negligible changes in the predicted ratios. 
To model the effects of extinction, the theoretical ratios were reddened using the \texttt{FM\_UNRED} routine in IDL, based on the extinction law of \citet{Fitzpatrick+99} and adopting a standard total-to-selective extinction law of R$_{V}$ = 3.1 (e.g., \citealt{Weingartne+01}). 
Figure~\ref{fig:extin} presents the observed strengths of the Brackett lines, normalized to the Br ${\gamma}$, together with the extinction-modified theoretical predictions. 
This comparison provides a direct estimate of the amount of extinction affecting the observed emission in G25.8+0.2.

To derive representative values of the visual extinction A$_{V}$ in G25.8+0.2, we employed two methods. First, we calculated the weighted mean of $A_V$ values derived from the Brackett line ratios. This yielded provisional estimates of A$_{V}$ = $9.3\pm1.1$ and $9.9\pm1.4$ mag for comp 2 and 3 in slit 1, and A$_{V}$ = $11.0\pm0.7$ mag for comp 2 of slit 2. As an alternative approach, we performed a $\chi^2$ minimization to fit the observed Brackett line ratios, treating A$_{V}$ as a free parameter varying over the range 4--22 mag. The best-fit values obtained from this method are A$_{V}$ = 10.9$^{+0.2}_{-0.2}$ and A$_{V}$ = 10.8$^{+0.5}_{-0.5}$ for comp 2 and 3 in slit 1, and A$_{V}$ = 11.1$^{+0.3}_{-0.3}$ mag for comp 2 of slit 2, with corresponding reduced ${\chi}^{2}$ values of 18.3, 7.2 and 4.3, respectively. The quoted uncertainties represent 90\% confidence intervals. 

The agreement between the two methods, within the quoted uncertainties, is good for slit 1 comp 3 and slit 2 comp 2, whereas for slit 1 comp 2, the values differ by ${\sim}$1.6 mag. This discrepancy for slit 1 comp 2 appears to arise from the deviation of the Br 14/Br ${\gamma}$ ratio from the Case B recombination prediction. As shown in Figure~\ref{fig:extin}, this line ratio exhibits a significant departure from the theoretical Case B curve. To assess its influence on the $\chi^2$ fit, we excluded the Br 14 line and repeated the analysis. The resulting best-fit extinction values are A$_{V}$ = 9.0$^{+0.4}_{-0.4}$ and 10.8$^{+0.5}_{-0.5}$ mag for slit 1 comp 2 and 3, with reduced $\chi^2$ values of 6.6 and 7.2, respectively. This confirms that the anomalous Br 14 ratio in slit 1 comp 2, which clearly deviates from the Case B prediction, is responsible for the inconsistency between the two methods.
 
Taking these considerations into account, we adopt A$_{V}$ = 9.0 ${\pm}$ 0.4 mag from the $\chi^2$ fitting for slit 1 comp 2, which agrees with the weighted mean (9.3 ± 1.1 mag). For slit 1 comp 3 and slit 2 comp 2, we adopt $\chi^2$-based values A$_{V}$ = 10.8 ${\pm}$ 0.5 mag and 11.1 ${\pm}$ 0.3, respectively, both consistent with the weighted means. The extinction-corrected Brackett line ratios, based on these values, are presented in Table~\ref{table:mecha}. 

For the shock-excited components (comp 1 in both slit positions), we applied extinction corrections based on the extinction values of the adjacent photoionized components (slit 1 comp 3 and slit 2 comp 2), under the assumption that the shock-excited and photoionized components are at the same distance. It is worth emphasizing that this study provides the first observational estimates of extinction toward G25.8+0.2. According to our analysis, the line-of-sight extinction toward the source lies in the range of 8.6--11.4 mag, with the value for slit 2 consistent with the upper end of this interval. A previous study (C18) inferred a broader extinction range of 10--16 mag, based on an assumed path length variation of 1.8 mag kpc$^{-1}$ and Galactic extinction models. This difference in extinction suggests that the actual extinction toward G25.8+0.2 may be somewhat lower than that inferred by C18 from Galactic-scale extinction models, implying that searches for ionizing stars under the assumption of 10--16 mag should be revisited to test whether photoionization alone can account for the observed ionization of G25.8+0.2.

\section{Discussion} \label{sec:dis}

\subsection{G25.8+0.2 in [Fe II] Emission and Radio Continuum} 
G25.8+0.2 appears as a semicircular arc with a radius of 3$\arcmin$, consisting of two extended filaments in [Fe II] emission (Figure 1). The two filaments -- hereafter referred to as the northern and southern filaments -- exhibit complex morphologies characterized by multiple overlapping small-scale loops. Our results reveal the presence of two distinct components of ionized gas toward the direction of the filaments. One component is prominent in [Fe II] emission but faint in Br ${\gamma}$ and He I lines (comp 1 in slits 1 and 2), whereas the other is bright in Br ${\gamma}$ and He I lines but shows little [Fe II] emission (comp 2 and 3 in slit 1, and comp 2 in slit 2). These components also differ in radial velocity: the [Fe II]-bright components have central velocities of 123 -- 138~\kms, while the Br $\gamma$-bright components range from 104 to 116 \kms. 

Strong [Fe II] lines are widely recognized as tracers of shocked gas, since Fe$^+$ is the dominant ionization stage of iron in radiative shocks and may also be enhanced by the destruction of dust grains \citep{Greenhouse+91, Nisini+05, Mouri+93, Mouri+00, Koo+15}. In contrast, [Fe II] emission is typically weak in H II regions because most iron exists in the higher ionization stage Fe$^{+2}$ and is largely depleted onto dust grains. The observed [Fe II]/Br ${\gamma}$ ratios of the Br $\gamma$-bright component are relatively low (0.12--0.33), comparable to a ratio typical of photoionized gas; for instance, [Fe II] 1.257/Pa$\beta$ ratios of 0.016--0.029 have been reported in Orion A \citep{walmsley+2000}, which translates to R([Fe II]/Br ${\gamma}$) of 0.069--0.12 using [Fe II] 1.644/Br${\gamma}\approx$ 4.3[Fe II] 1.257/Pa$\beta$ \citep{Koo+15}. In contrast, shock-ionized gas, such as that found in SNRs, exhibits ratios an order of magnitude or more higher (see \citealt{Koo+15}, and references therein). The extinction-corrected [Fe II]-bright component shows higher ratios, ranging from 1.33 to 1.97. In the two-dimensional spectra, the ratio reaches up to ${\sim}$4 in slit 1 and ${\sim}$6 in slit 2 (Figure~\ref{fig:2d}). These high ratios, observed in both slit positions of the northern and southern filaments, strongly suggest that the [Fe II]-bright component of the gas in these regions is shock-ionized. 

\begin{figure}
\begin{center}
\includegraphics[width=120mm]{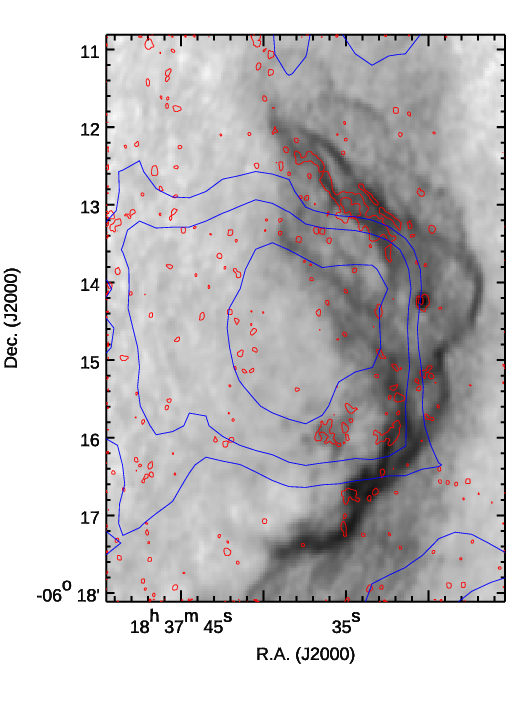} 
\end{center}
\caption{20 cm radio continuum image of G25.8+0.2 from the MAGPIS \citep{Helfand+06}. The image is displayed on a logarithmic scale.
Red contours indicate the continuum-subtracted [Fe II] emission shown in Figure 1.
Blue contours represent the ASCA soft X-ray emission, identical to the green contours in Figure 6 but smoothed for clarity.
The point-like source at R.A. = 18:37:30.42 and Dec. = -06:14:14.49 (J2000) is IRAS 18348$-$0616.} 
\label{fig:fe_radio}
\label{fig:zoomin}
\end{figure}

The velocities of the Br ${\gamma}$-bright component are consistent with those reported in previous radio recombination observations. In this region, a bright radio source showing striking morphological similarity to the IFO 96 is evident (Figure~\ref{fig:fe_2024},~\ref{fig:fe_radio}; see also C18). The radio source also forms a semicircular arc, although its structure is more intricate than that seen in [Fe II] emission. Both the northern and southern filaments consist of overlapping loop-like features, which are connected by an additional, smaller semicircular arc. At the center of this smaller arc lies a bright, compact source identified as a candidate ultracompact (UC) H II region (IRAS 18348-0616, \citealt{bronfman+1996}). In addition to the filamentary structures, diffuse emission is distributed across the region, with some parts appearing as faint filaments. The tilted ``A"-shaped structure (see Figure~\ref{fig:fe_radio}, R.A. =18:37:36, Dec. =-06:16:00) in the inner region of the main semicircular feature also shows a close morphological similarity between the two wavebands. The morphological correspondence observed suggests a physical link between the detected [Fe II] emission and at least a portion of the radio structure.

Several hydrogen recombination line observations have been carried out toward this source. \citet{Sewilo+04}, using a beam with a half-power beamwidth (HPBW) of 2\farcm56 ${\pm}$
0\farcm12 that encompassed both the northern and southern filaments as well as the central UC H II region, detected the H110${\alpha}$ line at a velocity of 112.1 ${\pm}$ 0.1 km s$^{-1}$. \citet{Quireza+06_recom} also detected the hydrogen recombination lines H91$\alpha$ and H92$\alpha$ at similar velocities of 111.44$\pm$0.17 and 111.35$\pm$1.89~\kms, respectively, along with helium recombination lines He91$\alpha$ and He92$\alpha$ at comparable velocities. These velocities are consistent with that of the Br $\gamma$-bright component of G25.8+0.2 (Table 2). The detection of H and He recombination lines indicates that the radio source is an H II region and that the Br $\gamma$-bright component detected in this work originates from ionized gas within that region.

In addition to the hydrogen and helium lines, \citet{Quireza+06_recom} also reported detections of carbon recombination lines C91$\alpha$ and C92$\alpha$, which exhibit significantly higher velocities of 124.52$\pm$0.96 \kms. This marked velocity offset relative to the hydrogen and helium lines suggests that the carbon lines do not originate in the same ionized region. Instead, their velocities are closer to those of the [Fe~II]-bright component of G25.8+0.2 (Table 2), suggesting a possible physical association between the [Fe~II]-emitting gas and the carbon recombination line-emitting region. Carbon recombination lines are generally thought to originate in photon-dominated regions (PDRs) or in partially ionized interfaces between ionized and molecular gas; thus, their velocity coincidence with the [Fe II]-bright component may indicate that both tracers originate in a transition zone between shock-ionized gas and adjacent molecular material.

The spectral properties of the radio continuum arc associated with G25.8+0.2 have been investigated in previous studies. \citet{DAmico+01} compared VLA DnC configuration images at 1.4 GHz and 4.8 GHz and derived a flat spectral index ($\alpha\sim 0$) for the brightest portion of the arc. They concluded that the emission likely originates from an H II region and suggested that nearby bright H II regions or an SNR expanding into a low-density environment could explain the absence of a detectable SNR in this region. We also derived the spectral index of the bright radio structure (i.e., G25.8+0.2) by comparing archival VLA interferometer data at 1.69, 2.97, and 4.71 GHz. The low-resolution  1.69--2.97 GHz data yields spectral indices of $\alpha\sim +0.09$ and +0.12 for the northern and southern structures, respectively, while the high-resolution 2.97--4.71 GHz gives $\alpha\sim +0.87$ and $-0.1$ for the northern and southern structures. These results are broadly consistent with the earlier measurements. 

However, \cite{Stein+21} reported a spectral index of $\alpha\sim -0.56$ between 2.7 GHz and 4.8 GHz, implying a possible synchrotron contribution. It should be noted that this spectral index was derived from a combination of single-dish and interferometric measurements probing different spatial scales, and a homogeneous radio study would therefore be valuable to further assess the nature of the non-thermal component. Our detection of strong [Fe II] emission further supports the possibility of non-thermal component in G25.8+0.2: the observed radio emission may not be purely thermal, but instead may represent a mixture of thermal (free-free) and non-thermal (synchrotron) emission. The coexistence of both mechanisms is plausible in complex environments such as massive star-forming regions, where photoionization from young massive stars and shocks from stellar winds or supernova activity can both contribute to the observed radio emission (see Section 4.3).

\subsection{Multi-wavelength View of the G26 system}

\begin{figure}
\center
\huge    
\includegraphics[width=155mm]{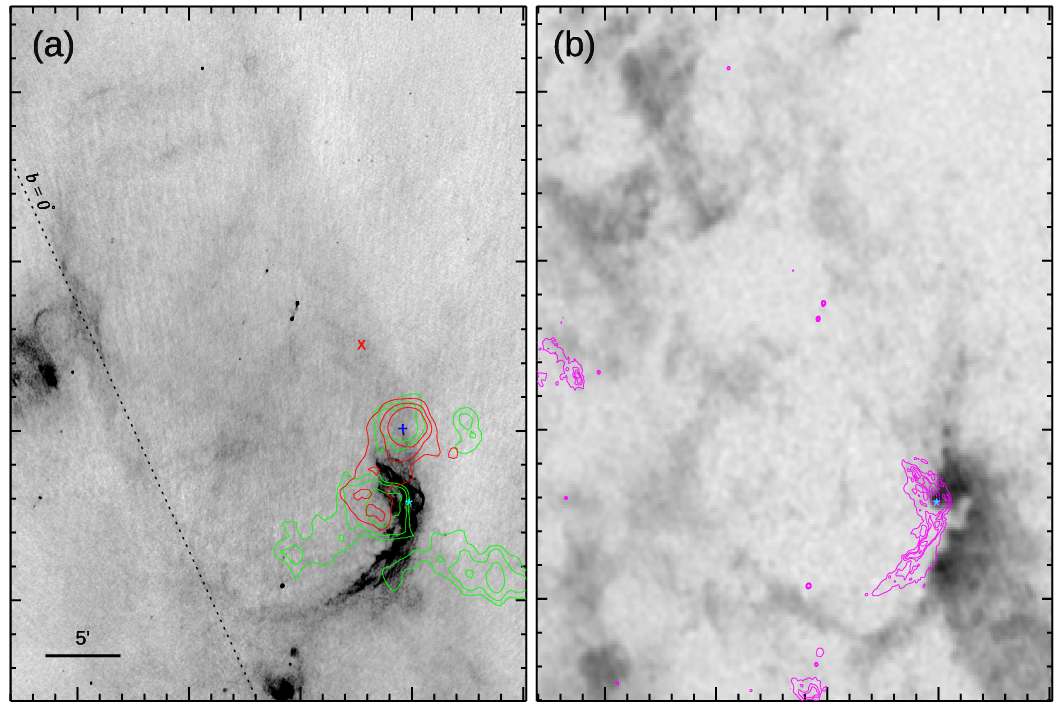}
\caption{
Zoomed-out view of the G $\sim$ 26$^{\circ}$ region in (a) 20 cm radio continuum and (b) $^{12}$CO \textit{J} = 1-0 emission. North is up and east is to the left.
The radio continuum image is from the MAGPIS, while the $^{12}$CO image is from the FUGIN survey \citep{Umemoto+17}, integrated over velocities from +106 to +115 km s$^{-1}$. The color scale in panel (b) is linear, spanning 0--195 K km s$^{-1}$.
The cyan asterisk marks the position of IRAS 18348$-$0616. In panel (a), green and red contours denote ASCA soft and hard X-ray emission, respectively. The blue plus symbol indicates the pulsar candidate AX J1837.5$-$0610, while the red cross marks PSR J1837$-$0604. The dotted line represents the Galactic midplane at \textit{b} = 0$^{\circ}$. Magenta contours in panel (b) show the radio continuum emission associated with the G25.8+0.2 structure shown in panel (a). }
\label{fig:out}
\end{figure}

Figure~\ref{fig:out} shows the 20-cm radio continuum image covering a wide field around  G25.8+0.2. A prominent feature in this image is a large U-shaped structure with an angular size of $\sim 20'{\times} 38'$, oriented at a position angle of roughly 25$^\circ$ (measured from north to east). G25.8+0.2 corresponds to the southwestern ``cap" of this structure. 
The lower portion of the U-shaped radio feature was first noted by \citet{DAmico+01}, and its potential association with G25.8+0.2 was later discussed in detail by C18. 

In that study, C18 found that the southwestern radio arc is in direct contact with a molecular cloud and interpreted the arc as a photoionized layer produced by massive stars located within the larger U-shaped structure. Supporting this interpretation, the molecular cloud exhibits enhanced CO emission along the radio arc, indicating a physical association between the ionized gas and the molecular material. This is illustrated in the $^{12}$CO map shown in Figure~\ref{fig:out}, integrated over the velocity range of +106 to 115 km s$^{-1}$, which corresponds to the systemic velocity of the region.
Note that this velocity range is consistent with that of the Br $\gamma$-bright component detected in this work.
This agreement provides strong evidence that the molecular cloud, the [Fe II] emission, and the U-shaped radio loop--including G25.8+0.2--are all physically associated, forming a coherent structure that we refer to as the G26 system. 

Figure~\ref{fig:out} also reveals the presence of diffuse molecular emission extending over a wide area matching the full spatial extent of the U-shaped radio structure. Additionally, an H I void was previously reported by C18 (see their Figure 4), encompassing the G26 system and sharing a comparable systemic velocity range of  +109.8 -- 118.1 km s$^{-1}$. Taken together, the presence of the H I void, the spatial correspondence between the molecular and ionized gas, and the consistent kinematic signatures all point to the G26 system being shaped or influenced by the energetic output—such as stellar winds or supernova explosions—from massive stars (C18). Further discussion of this feedback scenario is presented in Section~\ref{sec:feedback}.

In X-rays, several sources are seen around G25.8+0.2 (Figure~\ref{fig:out}). Most notably,
there is an extended soft X-ray  
(0.7 -- 2 keV) source to the east of G25.8+0.2, inside the U-shaped shell. This emission has a cometary morphology with an overall extent of $\sim 10'$. Its bright head, measuring $5'$ in diameter (or 9.5 pc at a distance of 6.5 kpc), 
seems to be enveloped by the arc-like structure that is bright both in radio and [Fe II] emission. The zoomed-in view of the soft X-ray, [Fe II], and radio continuum (Figure~\ref{fig:zoomin}) shows that the [Fe II] emission emerges where the soft X-ray emission is in close contact with the radio shell. \citet{Sakamoto+01} reported that the soft X-ray spectrum of this feature can be fitted with an optically thin thermal plasma model and suggested that it is likely a SNR (hereafter the G25.8+0.2 SNR candidate). 
The centrally brightened X-ray morphology is reminiscent of mixed-morphology SNRs (MMSRs), a class of remnants interacting with molecular clouds (\citealt{Rho+1998}; \citealt{Pannuti+2014}). Our detection of [Fe II] filaments with high [Fe II]/Br $\gamma$ ratios supports the SNR origin.

There are two other soft X-ray sources, one to the north of G25.8+0.2 and the other outside the radio structure to the southwest (see Figure~\ref{fig:out}). In the center of the former, there is a compact hard X-ray source AX J1837.5$-$0610, spatially coincident with an Energetic Gamma-Ray Experiment Telescope (EGRET) source (3EG J1837-0606, \citealt{Sakamoto+01}). 
\citet{Sakamoto+01} reported that
its hard X-ray spectrum is flat, and suggested
that it is probably a pulsar wind nebula. 
They further suggested an association of this hard X-ray source and a nearby molecular cloud based on the presence of broad CO line emission (see also \citealt{oka1999}). 
However, the density of the hydrogen column that absorbs the hard X-ray source obtained by \citet{Roberts+01}, is N$_{\rm H}$ = 0.58$^{+1.17}_{-0.58}$ ${\times}$ 10$^{22}$ cm$^{-2}$, which is lower than the column densities we infer toward G25.8+0.2. Our measurement span N$_{\rm H}$ ${\approx}$ 1.68$^{+0.08}_{-0.07}$ ${\times}$ 10$^{22}$ to 2.08$^{+0.05}_{-0.06}$ ${\times}$ 10$^{22}$~cm$^{-2}$ and exhibit an increasing trend from north to south. Given the large uncertainty in the \citet{Roberts+01} estimate and the fact that the hard X-ray source lies closest to our northern position, the upper end of their $N_{\mathrm{H}}$ interval may still be consistent with our measurements. Therefore, while a physical association between these X-ray sources and the [Fe~II] emission cannot be ruled out, it cannot be firmly established with the current data.

Two pulsars, PSR J1837-0604 and AX J1837.5-0610, are located in the vicinity of G25.8+0.2 (Figure~\ref{fig:out}). The estimated distance to PSR~J1837$-$0604, 6.41~kpc \citep{Mouri+05}, is in good agreement with that of the G26 system. Its characteristic spin-down age is $3.4\times 10^4$ yr \citep{DAmico+01}. Assuming an angular separation of $\sim10^{\prime}$ from the center of the G25.8+0.2 SNR candidate, the implied transverse velocity is ${\sim}$ 550 km s$^{-1}$, which is somewhat higher than the typical pulsar velocity but still within the observed range of high-velocity pulsars. Given the consistency in distance estimates between the pulsar and the G25.8+0.2 SNR candidate, a physical association between PSR~J1837$-$0604 and the SNR candidate remains plausible.

The pulsar candidate, AX J1837.5-0610, was identified using the same ASCA X-ray data analyzed by \citet{Sakamoto+01}, together with additional \textit{BeppoSAX} observations, and has been proposed as a radio-quiet $\gamma$-ray pulsar candidate with a characteristic age of $\sim1.5 \times 10^{3}$~yr \citep{Lin+08}. This source is located $\sim5^{\prime}$ from the center of the G25.8+0.2 SNR candidate, corresponding to a projected distance of $\sim9.5$~pc at 6.5~kpc.
If AX~J1837.5$-$0610 originated near the center of G25.8+0.2, a transverse velocity of $\sim6000$~km~s$^{-1}$ would be required to reach its current position within its characteristic age, which far exceeds the velocities observed for even the fastest known pulsars. Therefore, a physical association between AX~J1837.5$-$0610 and the G25.8+0.2 SNR candidate appears implausible. %

\subsection{Implications on the G26 Complex} \label{sec:feedback}

The G26 system is a complex and active star-forming region located in the inner Galaxy at a distance of approximately 6.5$\pm$1.0 kpc (C18). It is characterized by a large U-shaped radio continuum loop, with a prominent radio arc (G25.8+0.2) forming its southwestern cap. The full extent of the radio loop is $\sim 15’\times 30’$, corresponding to a physical size of $\sim 28$ pc $\times$ 57 pc, while the bright arc spans $\sim 20'$, or about 38 pc in physical scale. Toward G25.8+0.2, hydrogen radio recombination lines have been detected, suggesting that it is a H II region \citep{Sewilo+04, Quireza+06}. The flat radio continuum spectrum of G25.8+0.2 (\citealt{DAmico+01}; this study) also supports the H II region origin for the source. C18 carried out a comprehensive multiwavelength study of the G26 system and found that G25.8+0.2 is associated with an extended molecular cloud complex that appears elongated along the curvature of G25.8+0.2. The molecular cloud is composed of multiple dense clumps, several of which are likely active or potential sites of ongoing star formation. C18 proposed that the region's morphology and energetics are primarily shaped by feedback from massive OB-type stars via ionizing radiation, stellar winds, and SN explosions. They identified numerous candidate massive stars inside the U-shaped radio shell, including a Wolf-Rayet star, a red supergiant (RSG), and 37 2MASS sources with near-infrared colors consistent with O-type stars. These findings suggest recent and ongoing massive star formation. They further suggested that there might have been several SN explosions, although there are no catalogued SNRs within this region.

An extended soft X-ray source was detected in the vicinity of G25.8+0.2 by \citet{Sakamoto+01} during their investigation of a nearby EGRET gamma ray source. This X-ray source, which we refer to as the G25.8+0.2 SNR candidate, spatially fills the interior of the radio continuum arc G25.8+0.2. \citet{Sakamoto+01} reported that the spectrum of the source is well described by optically thin thermal emission from hot plasma, and suggested a SNR origin. Although the presence of this X-ray source was not discussed in the multiwavelength analysis by C18, it is potentially a remnant of a relatively recent SN event within the G26 system. Our detection of near-infrared [Fe II] emission from a shocked gas around the X-ray source lends further support to the SNR interpretation. The [Fe II] lines are often prominent in radiative shocks driven into dense material, consistent with what would be expected from an SNR interacting with its ambient medium.

The angular extent of the X-ray source corresponds to a physical radius of approximately 9.5 pc at the adopted distance of 6.5 kpc. For an SNR in the adiabatic (Sedov-Taylor) phase, its radius evolves as 
\begin{equation} 
R=12.5 E_{SN, 51}^{0.2} n_0^{-0.2} t_4^{0.4}~{\rm pc} 
\label{eq:sedov}
\end{equation}
where $E_{SN,51}$ is the explosion energy normalized to $10^{51}$~erg, $n_0$ is the ambient density in cm$^{-3}$, and $t_4$ is the age in $10^4$~yr. Adopting canonical values of $E_{SN}=10^{51}$~erg and $n_0=1$~cm$^{-3}$, the observed radius implies an age of $\sim 5.0\times 10^3$~yr, suggesting that G25.8+0.2 would be a relatively young to middle-aged SNR. Meanwhile, this age is substantially younger than the characteristic spin-down age of the nearby radio pulsar PSR J1837-0604 ($\tau\approx 34,000$ yr), which seem to suggest that they are unrelated. 

However, the morphology of G25.8+0.2 is inconsistent with that of a classical Sedov SNR: rather than exhibiting a limb-brightened shell, the X-ray emission is centrally brightened. This morphology is characteristic of MMSNRs \citep{Rho+1998}.
Mixed-morphology SNRs are commonly interpreted as remnants evolving in a highly inhomogeneous or clumpy ambient medium, such as a molecular cloud environment. In this scenario, dense clouds embedded within hot, X-ray-emitting plasma gradually evaporate, increasing the interior density and producing centrally filled thermal X-ray emission \citep{White+1991, Chevalier1999, 
Slavin+2017}. The expansion of the MMSNRs 
follows the same power law as for the Sedov SNRs, i.e., $R\propto t^{0.4}$ \citep{White+1991,Slavin+2017}.
Adopting a mean ambient density appropriate for a clumpy molecular environments, for example, $n_0\sim 10^{-3}$, Equation (\ref{eq:sedov}) yields an age of $t\sim 1.6\times 10^4$~yr. Modestly higher densities or a slightly lower explosion energy would increase the inferred age, bringing it into closer agreement with the characteristic age of PSR~J1837$-$0604.
Therefore, a mixed-morphology scenario allows for a significantly older remnant and does not rule out a physical association between G25.8+0.2 and the pulsar. Further studies are required to clarify the nature of G25.8+0.2 and its possible connection with PSR~J1837$-$0604.

The total 1.4 GHz flux density of the bright arc-shaped radio continuum source G25.8+0.2 (or G25.8700+0.1350 in C18) has been estimated as $S_{1420} = 12 \pm 1$ Jy, which implies a required number of ionizing photon rate of $N_{UV}=(5.0\pm 1.6) \times 10^{49}$ s$^{-1}$ to maintain the region ionized (C18). As noted by C18, this estimate represents a lower limit because the effect of the dust was not taken into account. They further suggested that such a photon budget could be supplied by several O type stars, like three O5V, 12 O7V or 63 O9V stars based on stellar atmosphere models by \citet{Martins+05}. Among the 37 2MASS sources identified as O-star candidates in the region, C18 concluded that many are likely physically associated with G25.8+0.2 and could plausibly be its ionizing sources.

We can constrain the spectral type of the ionizing stars using the He I 2.058 ${\mu}$m$/$Br ${\gamma}$ ratio derived in this work. This ratio, henceforth $R(\mathrm{He\,I}/\mathrm{Br}\,\gamma)$, is known to be sensitive to the effective temperature ($T_{eff}$) of the ionizing radiation field \citep[e.g.,][]{Shields+93,Lumsden+01,Lumsden+01_2,Lumsden+03}. For this purpose, we selected the photoionized components; comp 2, 3 of slit 1 and comp 2 of slit 2 (see Figure~\ref{fig:2d}). We compared the observed $R(\mathrm{He\,I}/\mathrm{Br}\,\gamma)$ values  with the photoionization model grids from \citet{Shields+93}, assuming a hydrogen density of  n$_{\rm H}$=10$^{2}$ cm$^{-3}$ and a filling factor ${\epsilon}$ = 0.01. These assumptions are consistent with the earlier adopted physical parameters of G25.8+0.2 (\citealt{Cichowolski+18}; n$_{e}$= 45.7 cm$^{-3}$, $\epsilon=0.08$). It is important to note that the He I 2.058 ${\mu}$m line is sensitive to several factors beyond just $T_{eff}$, including nebular geometry, optical depth effects and electron density \citep{Shields+93}. However, the line ratio remains a useful diagnostic in constraining the spectral type. 

To translate the inferred $T_{eff}$ into spectral types, we adopted the line-blanketed stellar atmosphere models of O dwarfs from \citet{Martins+02}. For comp 2 of slit 1 and slit 2, the observed line ratios (0.50${\pm}$0.2 and 0.59${\pm}$0.2) are consistent with ionizing stars of spectral type between O6.5 and O3 (38,000--49,000 K). Including comp 3 of slit 1, which has the lowest $R(\mathrm{He\,I}/\mathrm{Br}\,\gamma)$ ratio among the selected regions (0.36${\pm}$0.4), yields a similar constraint, indicating that the earliest ionizing source present is unlikely to be later than O7 (37,000 K). Thus, we conclude that the dominant ionizing stars in G25.8+0.2 must be of spectral type earlier than O7. To account for the required total ionizing photon budget derived from the radio continuum ($N_{UV}=(5.0\pm 1.6) \times 10^{49}$ s$^{-1}$; C18), this implies that at least three O3V stars or twelve O7V stars are necessary to sustain the observed ionization. Future spectral typing and radial velocity studies of O star candidates, taking into account the observationally constrained extinction range obtained in this study, are needed to confirm the true ionizing agents.

In the context of Galactic star formation, the G26 system can be compared to other large-scale, actively star-forming regions hosting massive stellar populations, such as the Carina Nebula. \cite{Rahman+2013} conducted a systematic search for massive young clusters and OB associations associated with the most luminous free-free emission complexes in the Milky Way, identifying 22 star-forming complexes with angular sizes of several to tens of arcminutes and stellar masses in the range $10^{2.3}$-$10^{5} M_{\odot}$. In this study, the G26 system—cataloged as SFC~12 with an estimated star-forming complex mass of $\ge10^{4.3}M_{\odot}$ (Table~3)—is comparable in scale to massive, active star-forming regions such as the Carina Nebula (${\textgreater}$$10^{4.3}M_{\odot}$), placing them among the seven most massive complexes. As emphasized by \cite{Rahman+2013}, such regions dominate the energetic feedback in the Galactic disk through photoionization, stellar winds, and supernova explosions, driving the formation of superbubbles and facilitating disk-halo interactions. 
The coexistence of photoionized gas, shock-excited [Fe~II] emission, molecular material, and large-scale radio structures in G26 suggests that it captures multiple phases of massive-star feedback operating simultaneously.

\section{Summary} 

G25.8+0.2 is an extended, arc-like radio structure. It is spatially correlated with IFO 96, which was serendipitously discovered in the UWIFE [Fe II] 1.644 ${\mu}$m imaging survey. It is located within the G26 complex, a large star-forming region in the inner Galaxy characterized by a large ($\sim 15'\times 30'$ or 28 pc $\times$ 57 pc at the distance of 6.5 kpc) U-shaped radio continuum shell and associated molecular material. Previous studies have primarily interpreted G25.8+0.2 as an H II region, although the presence of soft X-ray emission partially filling the radio shell has suggested a more complex origin.
To investigate its physical nature, we carried out high-resolution near-infrared \textit{H}- and \textit{K}-band spectroscopic observations using IGRINS at two slit positions sampling the bright [Fe II] filaments. 
The spectra reveal two distinct velocity components at both slit positions, characterized by markedly different [Fe II]/Br $\gamma$ ratios. The high-velocity components exhibit elevated ratios indicative of shock excitation, suggesting an association with a supernova remnant, while the low-velocity components are consistent with photoionized gas associated with an H II region. 
By combining these spectroscopic results with radio, X-ray, and molecular-line data, we explore the nature of G25.8+0.2 and its association with the G26 star-forming complex. The main results are summarized as follows.

\begin{enumerate} 

\item
At both slit positions, the [Fe II] emission exhibits two distinct velocity components. The high-velocity components show strong [Fe II] emission with elevated [Fe II]/Br $\gamma$ ratios, indicative of shock excitation, whereas the low-velocity components are dominated by hydrogen and helium recombination lines with weak [Fe II] emission, consistent with photoionized gas. The [Fe II]-bright components have velocities of $\sim$20--30 km s$^{-1}$, suggesting an origin in shocked circumstellar or interstellar material rather than in fast-moving SN ejecta.

\item
Using Brackett line ratios, we derived the first observational constraints on the extinction toward G25.8+0.2, obtaining $A_V \simeq 8.6$--11.4 mag. This range is both narrower and systematically lower than previous estimates based on Galactic extinction models.

\item
The [Fe II] filaments exhibit a strong morphological similarity with a portion of the bright radio shell within the G26 system and appear to be in contact with adjacent molecular clouds. They partially enclose centrally filled soft X-ray emission. This configuration—thermal X-ray emission surrounded by a radio shell and [Fe II] filaments—is characteristic of a supernova remnant interacting with dense ambient material. While radio spectral indices alone are inconclusive due to observational limitations, the combined infrared, X-ray, and radio evidence supports the presence of recent supernova activity within the G26 system.

\item
From the observed He I 2.058 ${\mu}$m/Br $\gamma$ ratio, we infer that the dominant ionizing stars in G25.8+0.2 are earlier than spectral type O7. 

\end{enumerate}

\clearpage

\begin{table}[ht]
\caption{Summary of IGRINS Spectroscopic Observations} 
\begin{center}
\begin{tabular}{@{}lccccclll} \toprule
\hline      
\colhead{} & \colhead{Date} & \phantom{a} & \colhead{Slit name} & \phantom{a} & \colhead{Slit position}  &  \phantom{a} &  \colhead{t$_{\rm exp}$$^{a}$} & \colhead{P.A.}  \\  
\colhead{} & \colhead{(UT)} &  \colhead{}      & \colhead{}            &  \colhead{}     & \colhead{(${\alpha}$, ${\delta}$ J2000)}  &  \colhead{}  & \colhead{(s)}  & \colhead{(deg)} \\  
\toprule
 & \multirow{2}{*}{2015-08-06} &  & Slit 1 &  & 18:37:32.62 \textminus06:13:17.4 & &  340${\times}$2 & 130  \\ 
 &  &  & Slit 2 &  & 18:37:32.38 \textminus06:15:56.5 & & 340${\times}$2 & 130  \\   
\bottomrule
\end{tabular}
\end{center}
\tablenotetext{a}{The total on-source exposure time.} 
\label{table:log}
\end{table}

\clearpage

\newcommand{\ra}[1]{\renewcommand{\arraystretch}{#1}}

\renewcommand{\tabcolsep}{3pt}

\movetabledown=2.7in
\begin{deluxetable}{@{}llccccccccccccccccc}
\tabletypesize{\scriptsize}
\tablecaption{Detected Lines and Their Properties}
\tablehead{
\phantom{a} & \multicolumn{2}{c}{Line} & \phantom{a} &  \multicolumn{14}{c}{Slit 1} &   \\
\cline{1-3}\cline{5-19}
\colhead{} & \multirow{2}{*}{Transition} & \multirow{2}{*}{$\lambda_{vac}$} & \colhead{} & \multicolumn{4}{c}{Comp 1} & \colhead{} & \multicolumn{4}{c}{Comp 2} & \colhead{} & \multicolumn{4}{c}{Comp 3} & \colhead{} \\
\cmidrule{5-8} \cmidrule{10-13} \cmidrule{15-19} 
& &  & & $v_{LSR}$ & $\Delta v$ & F$_{obs}^{a}$ & F$_{dered}^{a}$ & & $v_{LSR}$ & $\Delta v$ & F$_{obs}$ & F$_{dered}$ & & $v_{LSR}$ & $\Delta v$ & F$_{obs}$ & F$_{dered}$  & \\
& (1) & (2) & & (3) & (4) & (5) & (6) & &  (3) & (4) & (5) & (6) & & (3) & (4) & (5) & (6) &
}
\startdata     
 & \textbf{[Fe II]} & \textbf{1.6440} & & 127.6 & 17.2 & \textbf{1.15} & \textbf{1.97} & & 120.2 & 38.9 & \textbf{0.15} & \textbf{0.24} & &  103.2 & 19.7   & \textbf{0.19} & \textbf{0.33} & \\
 &                  &                 & & (0.3) & (0.7) &   (0.09)     & (0.16) & & (1.5)& (3.9) &   (0.02)      &  (0.03) & & (0.6) & (1.3) & (0.02)& (0.03) &  \\  
 & \textbf{Br ${\gamma}$} & \textbf{2.1661} & & 129.5 & 18.6 & 1.00$^{b}$ &  1.00 & & 115.8 & 29.6 &   1.00 &   1.00 & & 107.2 & 26.5 &   1.00  &   1.00 &  \\ 
 &                        &                 & & (0.5) & (0.8) & $\dotsm$ & $\dotsm$ & & (0.1) & (0.2) & $\dotsm$ & $\dotsm$ & & (0.3) & (0.5) & $\dotsm$ & $\dotsm$ &  \\  
 \cmidrule{2-19}                                  
& Br 14 & 1.5885 & & $\dotsm$ & 11.3  &  0.04  &  0.08 & & $\dotsm$ &  18.5 &  0.03 &  0.06 & & $\dotsm$ &  23.8 &  0.06 &  0.12  &   \\
&       &        & & $\dotsm$ & (6.2) & (0.03) & (0.06)& & $\dotsm$ & (3.9) & (0.01)& (0.02)& & $\dotsm$ & (4.5) & (0.01) & (0.03) &  \\ 
& Br 12 & 1.6412 & & $\dotsm$ &  17.6  & 0.05   &   0.09 & & $\dotsm$ &  29.4  & 0.10  &   0.15 & & $\dotsm$ &  36.6 &  0.13  &  0.22   &  \\
&       &        & & $\dotsm$ & (11.5) & (0.05) & (0.08) & & $\dotsm$ & (3.3) & (0.01) & (0.02) & & $\dotsm$ & (5.3) & (0.02) & (0.04)  &  \\ 
& Br 11 & 1.6811 & & $\dotsm$ &  14.6 &  0.08  &   0.14 & & $\dotsm$ &  34.9 &   0.20 &   0.30 & & $\dotsm$ &  24.7 &  0.12  &   0.19 &  \\
&       &        & & $\dotsm$ & (5.4) & (0.04) & (0.07) & & $\dotsm$ & (2.2) & (0.02) & (0.02) & & $\dotsm$ & (2.9) & (0.02) & (0.03) &  \\ 
& Br 10 & 1.7367 & & $\dotsm$ & 14.5  &   0.16 &  0.25  & & $\dotsm$ & 37.0  & 0.23   & 0.32   & & $\dotsm$ & 33.5  & 0.25   &  0.38 &  \\
&       &        & & $\dotsm$ & (2.3) & (0.04) & (0.06) & & $\dotsm$ & (1.9) & (0.02) & (0.02) & & $\dotsm$ & (2.1) & (0.02) & (0.03) &  \\ 
& He I  & 2.0587 & &    122.9 & 21.2  & 1.23   &   1.34 & & 116.0  &  24.3 &   0.47 &  0.50 & & 104.2  & 15.2  & 0.33  & 0.36 &  \\ 
&       &        & &    (1.0) & (1.7) & (0.13) & (0.14) & &(0.3) & (0.7) & (0.02) & (0.02) & & (0.5) & (1.6) & (0.04) & (0.04)  &  \\ 
\hline
\cmidrule{1-14}
\phantom{a} & \multicolumn{2}{c}{Line} & \phantom{a} & \multicolumn{9}{c}{Slit 2} & \\
\cline{1-3}\cline{5-14}  
\phantom{a} & \multirow{2}{*}{Transition} & \multirow{2}{*}{$\lambda_{vac}$} &\phantom{a} & \multicolumn{4}{c}{Comp 1} & \phantom{a} & \multicolumn{4}{c}{Comp 2}   \\
\cmidrule{5-8} \cmidrule{10-14}
& &  & & $v_{LSR}$ & $\Delta v$ & F$_{obs}^{a}$ & F$_{dered}^{a}$ & & $v_{LSR}$ & $\Delta v$ & F$_{obs}$ & F$_{dered}$ &  \\
& (1) & (2) & & (3) & (4) & (5) & (6) & &  (3) & (4) & (5) & (6) & \\
\cmidrule{1-14}
& \textbf{[Fe II]} & \textbf{1.6440} & & 139.0 & 19.7 & \textbf{0.76} & \textbf{1.33} &  & 111.9 & 18.5 & \textbf{0.07} & \textbf{0.12} & \\
&                  &                 & & (0.3) & (0.7) &      (0.04)  & (0.06)        &  & (0.8) & (1.8) &        (0.01) &      (0.01)  & \\ 
& \textbf{Br ${\gamma}$} & \textbf{2.1661} & & 138.2 & 24.1   & 1.00$^{b}$ & 1.00   & & 111.6 & 22.3 & 1.00   & 1.00  & \\ 
&                        &                 & & (0.2) & $\dotsm$ &   $\dotsm$   & $\dotsm$ & & (0.1) & (0.1) & $\dotsm$ & $\dotsm$ & \\ 
 \cmidrule{2-14}
& Br 14 & 1.5885 & & $\dotsm$ &  53.3  & 0.15  &  0.29 & & $\dotsm$  &  20.7 &   0.05 & 0.10 & \\
&       &        & & $\dotsm$ & (11.0) & (0.04) & (0.08)& & $\dotsm$ & (2.1) & (0.01) & (0.01) & \\
& Br 13 & 1.6114 & & $\dotsm$ &  41.6  & 0.20  &  0.16  & & $\dotsm$ &  19.6 & 0.07   &  0.12  & \\
&       &        & & $\dotsm$ & (21.6) & (0.06)& (0.10) & & $\dotsm$ & (2.7) & (0.01) & (0.02) & \\
& Br 12 & 1.6412 & & $\dotsm$ &  20.3 &  0.08  &   0.14 & & $\dotsm$ &  22.6 &   0.11 & 0.19 & \\
&       &        & & $\dotsm$ & (4.9) & (0.03) & (0.05) & & $\dotsm$ & (1.5) & (0.01) & (0.02) & \\ 
& Br 11 & 1.6811 & & $\dotsm$ & 33.2  &   0.18 &   0.30 & & $\dotsm$ & 22.5  & 0.15   &  0.24  & \\
&       &        & & $\dotsm$ & (6.3) & (0.04) & (0.07) & & $\dotsm$ & (1.1) & (0.01) & (0.02) & \\
& Br 10 & 1.7367 & & $\dotsm$ &  25.7 &   0.23 &   0.34 & & $\dotsm$ &  26.0 &   0.23 &  0.35 & \\
&       &        & & $\dotsm$ & (2.6) & (0.03) & (0.05) & & $\dotsm$ & (0.9) & (0.01) & (0.02) & \\ 
& He I  & 2.0587 & & 137.5 & 26.3  & 0.48   & 0.52   & & 110.7 &  18.3 &  0.55  &  0.59 & \\ 
&       &        & & (1.0) & (2.3) & (0.05) & (0.05) & & (0.2) & (0.4) & (0.01) & (0.02) & \\ 
 \cmidrule{1-14}
 \enddata
\tablenotetext{a}{F$_{obs}$ : observed flux, F$_{dered}$ : extinction-corrected flux. Fluxes are in 10$^{-15}$ erg s$^{-1}$ cm$^{-2}$.}
\tablenotetext{b}{Br ${\gamma}$ fluxes of comp 1, 2, and 3 of slit 1 are 3.98 ${\pm}$ 0.24, 13.12 ${\pm}$ 0.13, 13.14 ${\pm}$ 0.29 ${\times}$ 10$^{-15}$ erg s$^{-1}$ cm$^{-2}$. Br ${\gamma}$ fluxes of comp 1 and 2 of slit 2 are 13.64 ${\pm}$ 0.29 and 41.40 ${\pm}$ 0.26 ${\times}$ 10$^{-15}$ erg s$^{-1}$ cm$^{-2}$.}
\tablenotetext{*}{Br 13 in slit 1 was not detected.}
\tablenotetext{c}{Averaged FWHM of all detected lines weighted with 1 / ${\sigma}$$^{2}$, where ${\sigma}$ is statistical error.}
\tablenotetext{d}{Average radial velocity of all detected lines on each component at Local Standard-of-Rest frame. The uncertainty in parenthesis is propagated error of all detected lines for weighted mean value.}
\label{table:mecha}
\end{deluxetable}

\clearpage

\clearpage

\clearpage 

\section{Acknowledgments}
\begin{acknowledgments}
We are grateful to Hwihyun Kim, Kyle F Kaplan for their involving of IGRINS mini-queue observation. This work used The Immersion Grating Infrared Spectrometer (IGRINS) was developed under a collaboration between the University of Texas at Austin and the Korea Astronomy and Space Science Institute (KASI) with the financial support of the US National Science Foundation under grants AST-1229522, AST-1702267 and AST-1908892, McDonald Observatory of the University of Texas at Austin, the Korean GMT Project of KASI, the Mt. Cuba Astronomical Foundation and Gemini Observatory. This paper includes data taken at the McDonald Observatory of the University of Texas at Austin. 
B.-C. K. acknowledges support from the Basic Science Research 
Program through the NRF of Korea funded by the Ministry of
Science, ICT and Future Planning (RS-2023-00277370).
This publication makes use of data from FUGIN, the FOREST Unbiased Galactic plane Imaging survey with the Nobeyama 45 m telescope, a legacy project in the Nobeyama 45 m radio telescope. The 45 m radio telescope is operated by the Nobeyama Radio Observatory, a branch of the National Astronomical Observatory of Japan. 
\end{acknowledgments}

\vspace{5mm}
\facilities{UKIRT, Harlan J. Smith Telescope (IGRINS)}

\software{astropy \citep{2013A&A...558A..33A,2018AJ....156..123A},  
          IDL, IGRINS plp}

\bibliography{reference2}
\bibliographystyle{aasjournal}

\end{document}